\documentclass[12pt]{article} 
\pdfoutput=1
\usepackage[utf8]{inputenc}
\usepackage{graphicx}
\usepackage{amsmath}
\usepackage{amssymb}
\usepackage{units}
\usepackage{xcolor, adjustbox}
\usepackage[numbers, sort&compress, elide]{natbib}
\usepackage{jheppub}

\usepackage[normalem]{ulem}
\def\equationautorefname~#1\null{eq.\,(#1)\null}
\def\sectionautorefname~#1\null{sect.\,#1\null}
\def\subsectionautorefname~#1\null{sect.\,#1\null}
\def\figureautorefname~#1\null{fig.\,#1\null}
\def\appendixautorefname~#1\null{app.\,#1\null}

\makeatletter\g@addto@macro\bfseries{\boldmath}\makeatother

\newcommand{\bea}{\begin{eqnarray}}
\newcommand{\eea}{\end{eqnarray}}
\newcommand{\beq}{\begin{equation}}
\newcommand{\eeq}{\end{equation}}

\newcommand{\SON}{\text{SO}(N)}

\newcommand{\SONp}{\text{SO}(N+1)}
\newcommand{\GG}{\Pi}

\begin{document}

\title{Gegenbauer Goldstones}

\author[a]{Gauthier Durieux,}
\author[a]{Matthew McCullough,}
\author[a,b]{Ennio Salvioni}
\affiliation[a]{CERN, Theoretical Physics Department, Geneva 23 CH-1211, Switzerland}
\affiliation[b]{Universit\`a di Padova, Dipartimento di Fisica e Astronomia, Via Marzolo 8 I-35131, Italy}

\abstract{
We investigate radiatively stable classes of pseudo-Nambu-Goldstone boson (pNGB) potentials for approximate spontaneously broken $\SONp\to\SON$.
Using both the one-loop effective action and symmetry, it is shown that a Gegenbauer polynomial potential is radiatively stable, being effectively an `eigenfunction' from a radiative perspective.
In Gegenbauer pNGB models, one naturally and automatically obtains $v \propto f/n$, where $n\in 2\mathbb{Z}$ is the order of the Gegenbauer polynomial.
For a Gegenbauer Higgs boson, this breaks the usual correlation between Higgs coupling corrections and `$v/f$' tuning.
Based on this, we argue that to conclusively determine whether or not the Higgs is a composite pNGB in scenarios with up to $\mathcal{O}(10\%)$ fine-tuning will require going beyond both the Higgs coupling precision and heavy resonance mass reach of the High-Luminosity LHC.
}

\preprint{CERN-TH-2021-141}

\maketitle

\section{Introduction}
Nambu-Goldstone bosons (NGBs)~\cite{Goldstone:1961eq,Nambu:1960tm} are ubiquitous in theoretical physics, from phenomenological models to esoteric theories.
Their prevalence stems from the fact that a broad universality class of microscopic UV possibilities all lead, in the IR, to unique NGB theories whose leading structural features are dictated solely by the spontaneously broken global symmetries.
The specific details of the UV structure are only imprinted in operators which are increasingly irrelevant in the IR.
When the spontaneously broken global symmetries are also explicitly broken by some small parameter the structure of the IR theory depends only on the nature of the explicit breaking.
For non-linearly realised internal symmetries the consequent NGBs become pNGBs as they no longer parameterise a degenerate vacuum.

In principle, a pNGB potential could take any form, up to some non-trivial periodicity in terms of a decay constant `$f$' which is a consequence of the compactness of the underlying spontaneously broken symmetry.  One may wish to determine whether pNGB potentials can be understood in terms of some fundamental `building blocks' mapping the range of possible structures.  Here it is found that, at least for the symmetry-breaking pattern $\SONp\to\SON$, there is a sense in which there are fundamental building blocks of the scalar potential: Gegenbauer polynomials.

This conclusion is first derived from the perspective of renormalisation and radiative stability. Working within the IR theory one cannot ascertain the precise form UV quantum corrections would take within the full microscopic completion in which all IR properties are calculable in terms of the more fundamental UV parameters. However, the structure of UV corrections may be estimated from within the IR theory by studying the cutoff dependence of radiative corrections, assuming the physics at the cutoff does not introduce any additional explicit symmetry breaking.\footnote{Note that within the IR theory alone certain observables, such as pNGB masses, are incalculable and can only be fixed by measurement. However, by assumption the IR theory is the effective low-energy description of a theory in which physical properties are calculable in terms of more microscopic UV parameters, just as for the pions (IR) and the quark masses and QCD coupling (UV). Within this context, the na\"ive cutoff-dependence of IR quantities, subject to symmetries, provides a rough estimate as to the expected nature of the true UV-calculable contributions.}
One may thus ask what form of pNGB potential is insensitive to physics at the cutoff?
Here `insensitive' does not imply that UV corrections should be absent, but instead that they should not change the \emph{functional form} of the pNGB potential.
By examining the Coleman-Weinberg potential \cite{Coleman:1973jx} it is shown in \autoref{sec:CW} that Gegenbauer polynomials are, in this sense, eigenfunctions of the renormalisation at leading order in the small explicit symmetry-breaking parameter.
This is not the case for a generic scalar potential.
It suggests that the scope of possibilities for radiatively stable non-Abelian pNGB potentials and their associated phenomenology is much broader than the basic trigonometric functions found in typical models.

In \autoref{sec:spurion}, a deeper explanation based on spurions is presented.  If an explicit symmetry breaking spurion takes values in an irreducible representation (irrep) of $\SONp$ then at leading order in this spurion the generated pNGB potential will take the form of a Gegenbauer polynomial.
This is familiar from many areas of physics from scattering amplitudes to conformal field theories: the spin decomposition of functions (or operators) of momenta leads to Gegenbauer polynomials since they are the higher dimensional harmonic function counterparts of Legendre polynomials.  The difference is that in those cases the $\SONp$ symmetry is a spacetime symmetry and the variables are momenta or spacetime coordinates.  By contrast, here the $\SONp$ symmetry is an internal global symmetry and the coordinates are pNGB fields.  We may thus understand each individual Gegenbauer potential as effectively capturing a term in the multipolar expansion of the scalar potential, with the multipole living in the internal symmetry manifold.
In this way, the radiative stability found at one loop is also seen to persist to any loop order.  To construct the scalar potential the symmetry-breaking spurion may only be contracted with the nonlinear field parameterising the pNGBs, which transforms in the fundamental representation of $\SONp$, and the Kronecker-$\delta$.
The latter contractions vanish due to tracelessness of the symmetric spurion, being an irrep of $\SONp$.
Thus, at any radiative order and at leading order in the spurion, the only term that can be constructed in the scalar potential is that which corresponds to a Gegenbauer polynomial.\footnote{Going beyond the $\SONp\to\SON$ symmetry breaking pattern may modify both the appearance of Gegenbauer potentials and features of radiative stability.}

The above observations are interesting from a purely theoretical perspective and not widely appreciated. Some of them have been discussed for nonlinear sigma models in two dimensions in~\cite{Brezin:1976ap}, although with different implications.
There, Gegenbauer polynomials were already noted to be eigenfunctions of the renormalisation group flow, to arise from irrep spurions, and to therefore form a preferred basis for the potential decomposition.
However, these properties would become significantly more interesting if there were additional practical applications.
Here, we observe one feature which stands out in this regard.  For a Gegenbauer potential of  even order `$n$', with positive overall sign, the deepest minimum is at a Goldstone field magnitude `$v$' which scales as $v \propto f/n$.
Thus, for a large $n$, the pNGB expectation value is significantly smaller than the decay constant of spontaneous symmetry breaking. Importantly, the radiative stability of the potential guarantees that this fact survives radiative corrections and is technically natural.
In principle, this means that there is no obstruction to naturally finding $v\ll f$ for a pNGB.
It is well known that for Abelian pNGBs these features may readily be obtained through a $\mathcal{Z}_n$ symmetry arising from explicit breaking by an operator of charge $n$ but, to our knowledge, they have not been obtained before for non-Abelian pNGBs. 

In \autoref{sec:GBHiggs}, an application for these findings is presented in the context of a key question in fundamental physics, which is whether or not the Higgs boson could be a pNGB.
A composite pNGB Higgs, analogous to QCD pions, may explain a hierarchy between the electroweak scale $v$ and the scale of spontaneous symmetry breaking $f$ (and related new resonances).
However, it appears that in `generic' scenarios one must fine-tune to realise this hierarchy, at the level of $\Delta \approx 2 v^2/f^2$ (see e.g.\ \cite{Contino:2010rs,Bellazzini:2014yua,Panico:2015jxa} for discussions on this point).
While some constructions which challenge this expectation have been realised (see for instance~\cite{Harnik:2016koz,Galloway:2016fuo}), a clear picture has not yet emerged as to whether this required fine-tuning is a fundamental consequence of effective field theory (EFT) or if it is instead the result of some additional `minimality' assumption injected into the theory.
Based on the previous observations regarding Gegenbauer potentials, we argue for the latter.
As an explicit counterexample, we construct a `Gegenbauer Higgs' model in which the $v/f$ tuning is entirely absent thanks to a Gegenbauer potential and to its radiative stability.

Although the $v/f$ tuning is ameliorated in this Gegenbauer Higgs model, the fine-tuning stemming from the explicit symmetry breaking in the top-quark sector persists when reproducing the observed Higgs mass.\footnote{Note that constructions to ameliorate this tuning contribution have been found \cite{Csaki:2017cep}.}
For this reason we find that for $v/f \approx 1/8$, top-partner masses satisfying $M_{T} \approx 2$ TeV and single Higgs coupling modifications below $1\%$, a Gegenbauer Higgs model is fine-tuned at the $\sim 10\%$ level.
This is significantly less than standard pNGB Higgs models, where all explicit breaking arises from couplings to Standard Model (SM) fields in minimal spurion representations. Therefore, some relatively natural pNGB scenarios could remain inaccessible at the HL-LHC. Further experimental exploration to higher precision Higgs measurements and/or higher energy resonance searches will thus be necessary for a conclusive understanding of whether or not the Higgs is a pNGB.

\hyperref[sec:concl]{Section\,\ref{sec:concl}} contains some conclusions and a discussion of future avenues of investigation, which cover both unaddressed theoretical questions and the possibility of constructing Gegenbauer Higgs models that may also confront the apparent fine-tuning in the top-quark sector.

\section{Gegenbauer Goldstones}\label{sec:GG}
Consider Goldstone bosons from the spontaneous symmetry breaking $\SONp\to\SON$ with decay constant $f$.
The $N$ Goldstone bosons $\boldsymbol{\GG}$ sit in the fundamental representation of $\SON$ and are parameterised non-linearly as
\beq \label{eq:nonlinear_field}
\boldsymbol{\phi} =\frac{1}{\GG} \sin \frac{\GG}{f} \begin{pmatrix}
           \GG_{1} \\
           \GG_{2} \\
           \vdots \\
           \GG_{N} \\
          \GG \cot \frac{\GG}{f}
         \end{pmatrix}\;,
         \qquad\text{with }\quad \GG = \sqrt{\boldsymbol{\GG} \cdot \boldsymbol{\GG}}~,
\eeq
which, in accordance with the CCWZ procedure \cite{Coleman:1969sm,Callan:1969sn}, we treat as a fundamental of $\SONp$.  We will assume some small explicit symmetry breaking in the UV theory
such that the IR effective theory contains an $\SON$-invariant scalar potential
\beq
V = \epsilon M^2 f^2 G(\cos \GG/f) ~~,
\label{eq:potential}
\eeq	
where $G$ is some as-yet-undetermined function of $\cos \GG/f$, $\epsilon$ is a small dimensionless parameter, and $M$ is the typical UV mass scale.  As a result, the Goldstone bosons become pNGBs.

We wish to determine if there exists a class of scalar potentials $G$ whose \emph{functional form} is radiatively stable at $\mathcal{O}(\epsilon)$ against unknown UV corrections that respect the UV symmetry $\SONp$.  One way to estimate the scale of UV-corrections from within the IR theory is to use the one-loop Coleman-Weinberg (CW) potential, since hard cutoff renormalisation within the CW potential reveals the likely form of threshold corrections in matching from a given UV-completion to the IR theory.  Radiative stability requires that these threshold corrections must not alter the form of the scalar potential other than by a simple multiplicative factor. In other words, the structure of divergences should mimic the functional form of $G$, otherwise one would have to tune unknown deep UV-contributions against the threshold corrections in order to realise an IR effective potential $G$ whose functional form is significantly different from either of the former in isolation.

\subsection{Gegenbauers from Coleman-Weinberg}
\label{sec:CW}
Adopting a geometrical formulation, the general one-loop effective action for a set of scalar fields $\varphi^i$ with Lagrangian $\mathcal{L} = \tfrac{1}{2}g_{ij}(\varphi) \partial_\mu \varphi^i \partial^\mu \varphi^j - V(\varphi)$ may be determined following the formalism of \cite{Alonso:2015fsp}. This leads to the Coleman-Weinberg effective potential,
\begin{equation}
V_{\rm CW} = \frac{1}{2} \mathrm{Tr} \int \frac{d^4 p}{ (2\pi)^4} \log \left[p^2 + g^{-1} \left( \frac{\delta^2 V}{\delta \varphi^2} - \frac{\delta V}{\delta \varphi}\,\Gamma \right) \right]~~,
\end{equation}
with $p$ an Euclidean momentum,
and where the Christoffel symbols are
\beq
\Gamma^k_{\;\,ij} = \frac{1}{2} g^{kr} \left(\frac{\delta g_{ir}}{\delta \varphi^j} + \frac{\delta g_{jr}}{\delta \varphi^i} - \frac{\delta g_{ij}}{\delta \varphi^r} \right) ~~.
\eeq
Note that $g^{-1}(\delta^2V/\delta\varphi^2-\delta V/\delta\varphi \;\Gamma)$ is the Laplacian of $V$ in general $\varphi^i$ coordinates.

In our specific case, symmetry forces the potential to only depend on $\GG$ and the field metric is~\cite{Alonso:2015fsp}
\beq
g_{ij} = \begin{pmatrix} F(\GG)^2 \hat{g}_{ab} & 0\\0 & 1 \end{pmatrix}\,, \qquad F = \sin \frac{ \GG}{f}\;, 
\label{eq:fieldmetric}
\eeq
where $\hat{g}_{ab}$ with $a,b \in\{ 1, \ldots, N-1\}$ is a metric on the sphere $S^N$ spanned by the $\GG_i$ for fixed $\GG$. The quadratically divergent piece of the one-loop effective potential is
\beq
V_{\rm CW}^{\Lambda^2} = \frac{\Lambda^2}{32\pi^2}  \left( \frac{\delta^2 V}{\delta \GG^2} + (N - 1) \frac{1}{F} \frac{\delta{F}} {\delta \GG}\frac{\delta V}{\delta \GG} \right),
\label{eq:quadratic-cw}
\eeq
where $\Lambda$ is the hard UV-cutoff.
Thus, at the linear order in $\epsilon$, the quantum-corrected potential following from \autoref{eq:potential} is
\beq
V_Q = \epsilon M^2 \left( f^2  G + \frac{\Lambda^2 }{32 \pi^2} \Big(G''+(N-1) \cot\frac\GG{f} \, G' \Big) \right)
\eeq
where primes denote derivatives of $G(\cos \GG/f)$ with respect to $\GG/f$. We see from this one-loop correction that
a generic pNGB potential will not be radiatively stable against UV corrections.  On the contrary, for a scalar potential comprised of a function $G$ satisfying the strict relation
\beq \label{eq:mult_ren}
G''+(N-1) \cot\frac\GG{f}\, G' \propto G ~~,
\eeq
the one-loop quadratic divergences may be absorbed into a multiplicative counterterm at leading order in $\epsilon$. Gegenbauer polynomials $G^\lambda_n(\cos \GG/f)$ satisfy precisely this requirement. Specifically, they obey the second-order differential equation\footnote{Solutions for non-integer $n+\lambda-1/2$, as well as the second family of solutions obtained for integer values, are not analytic at certain field values and can therefore not be generated in an effective field theory where the only light particles are the Goldstone bosons we discuss.}
\beq \label{eq:Gegen_diff}
G_n^{\lambda\,\prime\prime} + 2 \lambda  \cot\frac\GG{f} G_n^{\lambda\,\prime} + n (n + 2\lambda) G^{\lambda}_n = 0 ~~,
\eeq
where $n\geq 0$ is an integer, so that \autoref{eq:mult_ren} is satisfied for $\lambda = (N-1)/2$.
As a consequence, Gegenbauer polynomials are uniquely identified as describing an infinite class of radiatively stable scalar potentials for pNGBs in the fundamental representation of $\SON$.
In other words, we find that the functional form of the scalar potential
\beq
V = \epsilon M^2 f^2\;  G^{(N-1)/2}_n (\cos \GG/f)
\label{eq:GGP}
\eeq
is radiatively stable against UV-physics at one loop and linear order in $\epsilon$.

Following the same approach that led to the linear contribution in \autoref{eq:quadratic-cw}, the quadratic contributions in $\epsilon$ that arise at the one-loop order and spoil the multiplicative renormalization of the potential are of the form
\begin{equation}
\left( \frac{\delta^2 V}{\delta \GG^2}\right)^2 + (N - 1) \left(\frac{1}{F} \frac{\delta{F}} {\delta \GG}\frac{\delta V}{\delta \GG} \right)^2\:.
\end{equation}
For such subleading terms in $\epsilon$ to remain negligible, one must require $\epsilon \:n^2\ll1$.
Given the differential equation in \autoref{eq:Gegen_diff}, the second derivative of a Gegenbauer potential indeed receives contributions scaling as $n^2$.

In the Abelian $N=1$ case, \autoref{eq:Gegen_diff} reduces to $G'' + n^2 G = 0$ whose solution is simply $\cos(n\GG/f)$ and is radiatively stable.
This is straightforward to understand in terms of symmetries, as this potential is invariant under a $\mathcal{Z}_n$ symmetry transforming $\GG\to \GG + 2\pi f/n$. The connection with symmetries for the non-Abelian case $N > 1$, which is far less obvious, is established in \autoref{sec:spurion}.

\begin{figure}\centering
\includegraphics[width=.5\textwidth]{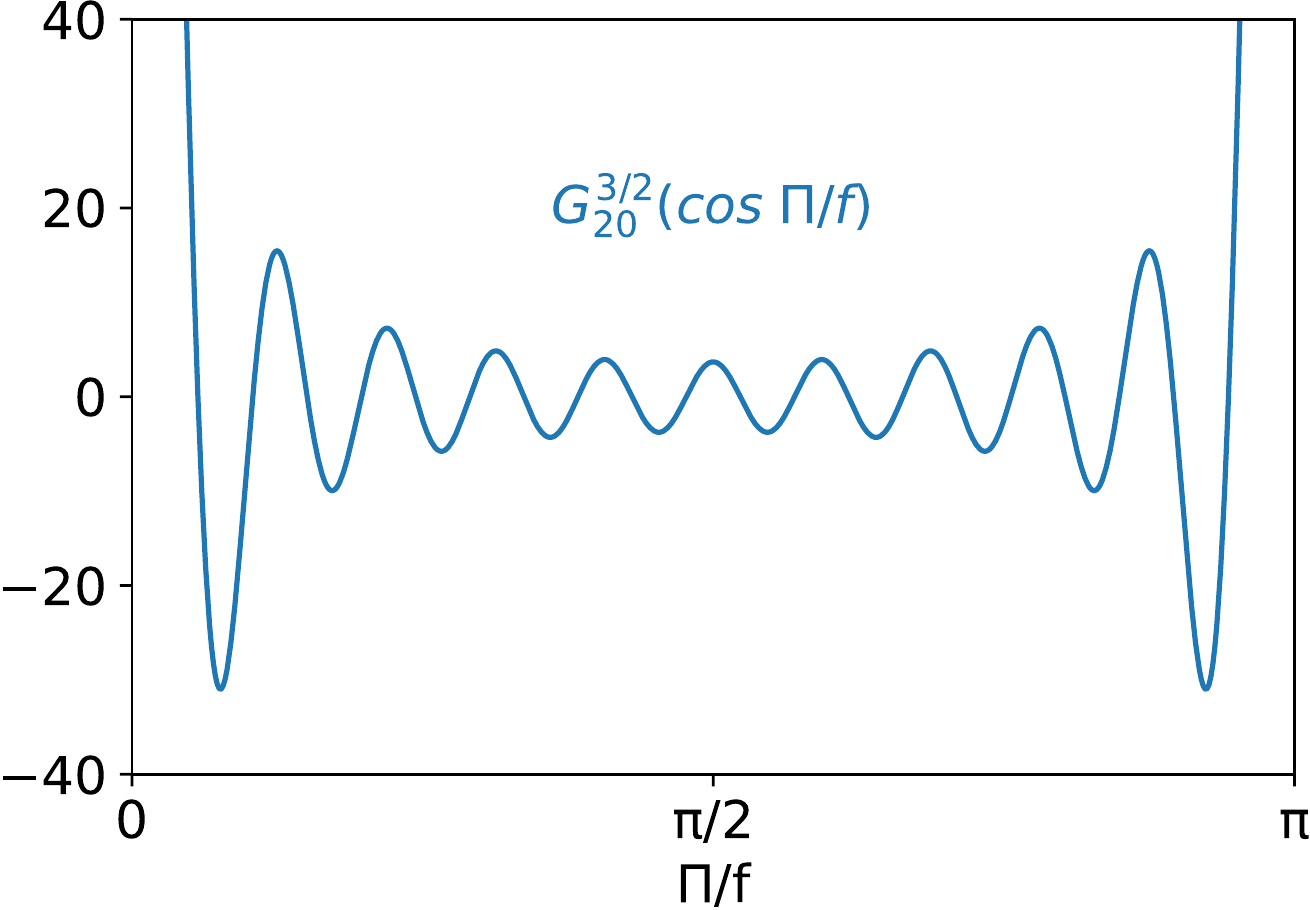}
\caption{For $\GG/f$ away from multiples of $\pi$ and $n$ large, the Gegenbauer polynomials $G^{(N-1)/2}_n(\cos\GG/f)$ approximately oscillate as $\cos n\Pi/f$.
For even $n$, their deepest minimum is the closest to $k\pi$ with $k\in \mathbb{Z}$.
For odd $n$, they take negative values at $(2k+1)\pi$ where the deepest minimum is then also located.
}
\label{fig:gegenbauer}
\end{figure}

Interestingly, there exists a regime in which $G^{(N-1)/2}_n(\cos\GG/f)$ oscillate approximately like $\cos(n\GG/f)$.
This can be seen graphically in \autoref{fig:gegenbauer}, obtained for $N=4$ and $n=20$.
More precisely, one can derive the following approximations (see \autoref{app:large-n}), up to overall normalisation factors
\begin{equation}
G^\lambda_n\left(\cos \frac{\GG}{f} \right)
	\;\xrightarrow{n\gg1}\; \frac{J_{\lambda-1/2}\left((n+\lambda) \frac{\GG}{f} \right)}{\GG^{\lambda-1/2}}
	\;\xrightarrow{\frac{\GG}{f}\gg\frac{1}{n}}\;  \frac{\cos\left((n+\lambda) \frac{\GG}{f}-\lambda\frac{\pi}{2}\right)}{\GG^\lambda}
	\,,
\label{eq:geg-approx}
\end{equation}
where $J_{\lambda-1/2}$ is the Bessel function of the first kind with order $\lambda-1/2$.
Thus for large $n$ and $\GG/f\gg 1/n$, we see that the potential becomes approximately periodic in $\GG$.
This is non-trivial and is, again, somewhat reminiscent of the potential for an Abelian pNGB in the case where an explicit symmetry-breaking spurion has large U$(1)$ charge.  Thus, while a purely periodic potential for a non-Abelian pNGB is not radiatively stable and thus not technically natural, the Gegenbauer polynomials do provide a potential which is radiatively stable and approximately periodic in $f/n$.  As we will see, this can have non-trivial consequences for phenomenological applications.

\subsection{Gegenbauers from Spurion Irreps}
\label{sec:spurion}
Taking a more symmetry-based approach, one can see that the radiative stability of the form of the potential in \autoref{eq:potential} would be guaranteed if it arose from the interaction of $\boldsymbol{\phi}$ with an explicit symmetry-breaking spurion $K$ sitting in a traceless symmetric irreducible representation of $\SONp$,
\beq \label{eq:spurion}
V = \epsilon M^2 f^2 K_n^{i_1 i_2 ... i_n}\; \phi^{i_1} ...\; \phi^{i_n}\,.
\eeq
Such a representation has dimension $(N+2n-1)(N+n-2)!/n!(N-1)!$, growing like $2n^{N-1}/(N-1)!$ for $n\gg N$.

The reason for stability is that radiative corrections at all scales will generally furnish every $\SONp$ invariant that can be constructed in the scalar potential out of the fields and parameters.
However, at $\mathcal{O}(\epsilon)$ the only structures we have to work with are $\boldsymbol{\phi}$, transforming as a fundamental, the $\SONp$-invariant Kronecker-$\delta$, and the spurion tensor $K_n^{i_1 i_2 ... i_n}$.
Hence, at leading order in $\epsilon$, the only non-vanishing term that can be constructed is that of \autoref{eq:spurion}.
Thus we conclude that the functional form of a pNGB potential generated from an explicit symmetry-breaking spurion in the traceless symmetric representation of $\SONp$ is radiatively stable against UV-corrections to all loop orders at $\mathcal{O}(\epsilon)$.
If the potential \autoref{eq:potential} could be constructed in this way, it would also be stable against radiative corrections to all loop orders at $\mathcal{O}(\epsilon)$.

A Gegenbauer polynomial does indeed arise from such a spurion.
This connection is well known (see e.g.~\cite{Brezin:1976ap,Chetyrkin:1980pr,Kazakov:1986mu}), however since we could not find a detailed explanation in the literature we present a short derivation here for completeness.
It rests on a parallel with the multipole expansion of axisymmetric potentials in general dimension.

Let us perform a Taylor expansion around $\boldsymbol{\tilde{\phi}}$ of the scalar function
\beq \label{eq:Taylor}
|t \boldsymbol{\phi} - \boldsymbol{\tilde{\phi}} |^{1-N} = \sum_{n=0}^\infty t^n  K_n^{i_1 i_2 ... i_n} (\boldsymbol{\tilde{\phi}} ) \phi^{i_1} \phi^{i_2} ... \phi^{i_n} ~~,
\eeq
where $t$ is a dimensionless parameter and we have now defined the $K_n^{i_1 i_2 ... i_n}$ tensor through\footnote{For illustration, an explicit construction of e.g.~$K_6$ is:
\begin{equation*}
\begin{aligned}
K_6^{i_1 i_2 i_3 i_4 i_5 i_6}(\boldsymbol{\tilde{\phi}})
&\sim \tilde{\phi}^{i_1}\tilde{\phi}^{i_2}\tilde{\phi}^{i_3}\tilde{\phi}^{i_4}\tilde{\phi}^{i_5}\tilde{\phi}^{i_6}
\\&- \tilde{\phi}^{i_1}\tilde{\phi}^{i_2}\tilde{\phi}^{i_3}\tilde{\phi}^{i_4} \delta^{i_5,i_6} /(N+9)+ \text{14 perm.}
\\&+ \tilde{\phi}^{i_1}\tilde{\phi}^{i_2}\delta^{i_3,i_4} \delta^{i_5,i_6} /(N+9)(N+7) + \text{44 perm.}
\\&- \delta^{i_1,i_2}\delta^{i_3,i_4} \delta^{i_5,i_6} /(N+9)(N+7)(N+5) + \text{14 perm.}
\:.
\end{aligned}
\end{equation*}}
\beq
K_n^{i_1 i_2 ... i_n} (\boldsymbol{\phi})= \frac{1}{n!}  \frac{ \partial^n \phi^{1-N}}{\partial \phi_{i_1} \partial \phi_{i_2} ... \partial  \phi_{i_n}} \,.
\label{eq:constrep}
\eeq
Clearly $K_n^{i_1 i_2 ... i_n} (\boldsymbol{\phi})$ forms an $n$-index representation of $\SONp$.
This representation is manifestly symmetric by construction and also traceless, since the $N+1$-dimensional Laplacian vanishes when evaluated away from the origin,
\beq
\sum_{i=1}^{N+1} \frac{\partial^2 \phi^{1-N}}{\partial \phi_{i}^2}
= \frac{1}{\phi^N} \frac{\partial}{\partial\phi}\left( \phi^N\frac{\partial\phi^{1-N} }{\partial\phi} \right)
= 0~~.
\eeq
Thus we have constructed the desired traceless symmmetric $n$-index representation of $\SONp$ from this higher-dimensional multipolar expansion (in field-space).  Now taking a constant value for $K_n^{i_1 i_2 ... i_n} (\boldsymbol{\tilde{\phi}})$ to explicitly break $\SONp$ this spurion will preserve the $\SON$ subgroup if we take $\boldsymbol{\tilde{\phi}}=(0,\ldots,0,1)^T$. This completes the construction of the desired class of symmetry-breaking spurions.

On the other hand, taking $\boldsymbol{\phi}$ as defined in \autoref{eq:nonlinear_field} we find
\beq
|t\boldsymbol{\phi} - \boldsymbol{\tilde{\phi}}| = \sqrt{1-2 t \cos \GG/f+t^2} \,,
\eeq
still for $\boldsymbol{\tilde{\phi}}=(0,\ldots,0,1)^T$.  The left-hand side of \autoref{eq:Taylor} can then also be recognised as the generating function of Gegenbauer polynomials
\beq
(1-2 t x + t^2)^{- \lambda} = \sum_{n=0}^{\infty} t^n G_n^{\lambda} (x) \,,
\eeq
for $\lambda=(N-1)/2$ and $x = \cos \Pi/f$. Identifying the two sides of \autoref{eq:Taylor} order-by-order in $t$, one therefore obtains
\beq
 K_n^{i_1 i_2 ... i_n} ( \boldsymbol{\tilde{\phi}})\; \phi^{i_1} \phi^{i_2} ... \phi^{i_n} = G_n^{(N-1)/2} (\cos \GG/f )~~.
\label{eq:tensor}
\eeq
We have hence shown that a $K_n^{i_1 i_2 ... i_n} (\boldsymbol{\tilde{\phi}})$ spurion irrep of $\SONp$ which explicitly breaks $\SONp\to \SON$ yields a Gegenbauer polynomial as scalar potential.
Thus the underlying reason for the radiative stability of the Gegenbauer polynomial pNGB potential found from the Coleman-Weinberg calculation is symmetry.
Furthermore, the spurion analysis reveals that at $\mathcal{O}(\epsilon)$ the form of a Gegenbauer pNGB potential will be radiatively stable at all loop orders, going beyond the Coleman-Weinberg calculation which only confirmed this radiative stability at one loop.

In general models, the presence of this symmetry-breaking spurion will correct the potential for the radial mode $\rho$ and hence the value of $f$.  However this is at $\mathcal{O}(\epsilon)$ and will thus be small if $\epsilon$ is small and will be particularly suppressed in strongly coupled models.  Furthermore, in a general theory there will be a tower of symmetry preserving operators which generate corrections to the radial mode potential $\propto \rho^p$ and thus a full knowledge of the UV theory is required to estimate the impact of the spurion on the radial mode potential.  We implicitly assume that the value of $f$ is the radial mode vacuum expectation value (vev) after including all of these effects.

\subsection{Gegenbauer Decomposition of a pNGB Potential}
It is thus established, from two complementary perspectives, that for pNGBs in the fundamental of $\SON$ the form of the scalar potential \autoref{eq:GGP} is radiatively stable and hence technically natural, for any $n$.
This is also suggestive that for these pNGBs the Gegenbauer polynomials form a preferred basis, from a quantum perspective, for the decomposition of any functional form of a pNGB potential.
The reason for this is that they also form a complete basis, orthogonal with respect to the weight $\sin^{N-2}(\GG/f)$. As a result any potential for  pNGBs in the fundamental of $\SON$ may be decomposed as
\beq
V = \epsilon M^2 f^2  \sum_{n=0}^{\infty} a_n G^{(N-1)/2}_n (\cos \GG/f) ~~.
\eeq
Each term is radiatively stable at all loop orders and $\mathcal{O}(\epsilon)$.  Divergent UV-corrections will treat the individual terms differently, by the factor $n(n+\lambda)$ from \autoref{eq:Gegen_diff} but will not mix them at this order.

This expansion is familiar for good reason.
In $D=3$ dimension electrostatics, for instance, one may decompose any axisymmetric charge distribution as an infinite sum of multipoles which are, from a symmetry perspective, $n$-index spurions in symmetric traceless representation of $\text{SO}(D=3)$.  This is the familiar decomposition in terms of Legendre polynomials, which are just special cases of the Gegenbauer polynomials for $D=3$.
Indeed, the same procedure in a greater number of spatial dimensions would result in an expansion in Gegenbauer polynomials.
The distinction here is that, in the electrostatics case, the polynomials arise as a result of explicit breaking of the \emph{spatial} symmetries $\text{SO}(D)  \to \text{SO}(D-1)$, whereas in the pNGB case they arise as a result of the explicit breaking of the \emph{internal} continuous symmetries $\SONp\to\SON$.  Nevertheless, the potential becomes an effective multipole expansion in this internal scalar field space.

\section{Gegenbauer Higgs}
\label{sec:GBHiggs}
Since the Gegenbauer potential for a pNGB can significantly alter the vacuum structure of a theory, it is natural to investigate potential areas of relevance in physics.
The Higgs boson is an obvious candidate application since, as yet, it remains to be understood whether it is a fundamental particle down to ultra short scales or if it is instead a composite pNGB.
The latter class of models has been explored in depth, but almost exclusively under minimal assumptions concerning the explicit symmetry-breaking terms, taken to originate only from the couplings of elementary fields to the composite sector and transforming in the few smallest representations of the global symmetry.
As we will now see, the introduction of a Gegenbauer potential for a composite pNGB Higgs can challenge basic expectations for the vacuum structure. 

We assume the Higgs doublet emerges among the $N$ fundamental NGBs of the $\SONp\to\SON$ spontaneous symmetry breaking. Henceforth we often focus on $N = 4$, corresponding to the minimal composite Higgs model (MCHM)~\cite{Agashe:2004rs}, but note that our discussion could be extended to non-minimal constructions with $N > 4$, where additional NGBs accompany the Higgs.
The kinetic term of the nonlinear sigma model reads
\begin{equation}
\mathcal{L}_2 = \frac{f^2}{2} D_\mu \boldsymbol{\phi}^T D^\mu \boldsymbol{\phi}~,
\end{equation}
where $\boldsymbol{\phi}$ was defined in \autoref{eq:nonlinear_field} and the covariant derivative gauges the SM electroweak group.
As a result, we identify the vev of the Higgs field as $f^2 \sin^2 \langle \GG \rangle/f = v^2 \approx (246\;\mathrm{GeV})^2$, and the Higgs boson coupling to SM gauge bosons is
\begin{equation}
c_{hVV}/c_{hVV}^{\rm SM} = \cos{\langle\Pi\rangle}/{f} = \sqrt{1 - v^2/f^2}~,
\end{equation}
one of the central predictions of this class of models.

Now we introduce explicit $\text{SO}(N+1)$ breaking in the form of a traceless symmetric $n$-index spurion irrep, which as shown in~\autoref{sec:GG} gives rise to a Gegenbauer Higgs potential $V(\Pi) = \epsilon M^2 f^2 G_n^{(N-1)/2} (\cos \GG/f)$. As seen in \autoref{fig:gegenbauer}, for positive coefficient $\epsilon$ and even $n$ the deepest minimum is the first, which for large $n$ is located at $\GG \ll f$.
In the large $n$ limit, its position may be obtained from the Bessel approximation as
\beq
\frac{\langle \GG \rangle}{f} \approx \frac{j_{\lambda+1/2,1}}{n+\lambda} \approx \frac{5.1}{n}~,
\eeq
where $j_{\lambda+1/2,1}$ denotes the first zero of the Bessel function of order $\lambda+1/2$. Its numerical value for $\lambda = (N-1)/2$ with $N = 4$ is $j_{2,1} \approx 5.1356$. Importantly, this global minimum is at a small non-zero value, scaling inversely with $n$. Thus Gegenbauer potentials provide a radiatively stable means by which to realise a Higgs vacuum expectation value much lower than the scale of spontaneous symmetry breaking in pNGB Higgs models.

\begin{figure}
\begin{center}
\includegraphics[width=.5\textwidth]{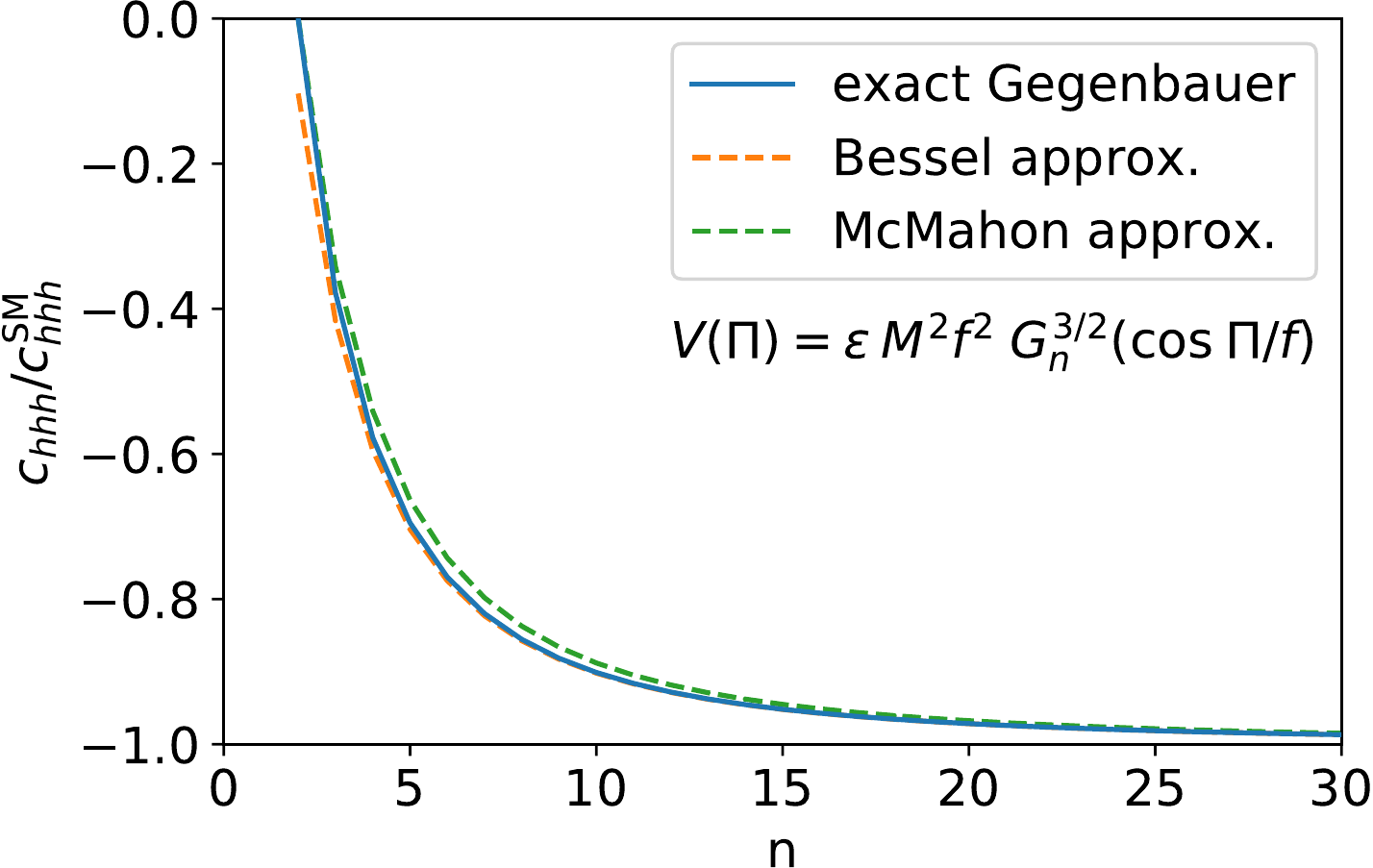}
\caption{Higgs trilinear self-coupling as a function of $n$, for a pure Gegenbauer potential $V(\GG)=\epsilon\: M^2 f^2\; G_n^{(N-1)/2}(\cos\Pi/f)$ with $N=4$.
In the large-$n$ limit, the location of the global minimum is approximated using the first Bessel zero ${\langle \GG \rangle}/{f} \approx {j_{\lambda+1/2,1}}/{(n+\lambda)}$.
The latter can further approximated by the first term of the McMahon formula  $j_{\lambda+1/2,1} \approx \pi (1+\lambda/2)$.
}
\label{fig:trilinear_behaviour}
\end{center}
\end{figure}

An additional novelty of Gegenbauer Higgs potentials concerns the trilinear Higgs self-coupling if one, for now, sets aside important contributions from the top and gauge sectors. Its value can be obtained by differentiating the potential with respect to $z = \Pi/f$ as $c_{hhh}/c_{hhh}^\text{SM} = [\sin z\; V'''/(3 V'')]_{ z\, =\, \langle \GG \rangle /f\,}$, where $c_{hhh}^\text{SM}=m_h^2/(2v)$. Taking an additional derivative of the Gegenbauer differential equation \autoref{eq:Gegen_diff} and setting the first derivative to zero, one finds that for a $G^{(N-1)/2}_n(\cos \GG/f)$ potential the correction is
\begin{equation}
\frac{c_{hhh}}{c_{hhh}^\text{SM}}
	=-\frac{N-1}{3}\; \cos \frac{\langle \GG \rangle}{f}
	\approx -\, \frac{N - 1}{3}~,
\label{eq:pure-geg-trilinear}
\end{equation}
where, in the last approximate equality, we have taken the large $n$ limit. We learn that for $N = 4$ the Higgs trilinear self-coupling with a purely Gegenbauer potential has {\it equal magnitude but opposite sign} compared to the SM, as illustrated in~\autoref{fig:trilinear_behaviour}.

While the above features are genuine novelties in the pNGB Higgs context, a complete model must also account for the explicit symmetry breaking originating from the couplings of the elementary degrees of freedom (transverse gauge fields and fermions) to the composite sector which, in turn, contribute radiatively to the Higgs potential. As the next step in the direction of a fully realistic model, we now include the dominant explicit breaking from the top-quark sector, which will have a significant impact on the Higgs self-coupling and fine-tuning in the model.

\subsection{Towards a Realistic Model}\label{sec:toward_realistic}
In pNGB Higgs models the dominant radiative contributions to the Higgs potential arise from the top quark and its composite partners, due to the large Yukawa interaction. The minimal magnitude of this potential is fixed by the lower bound on the top partner masses from direct searches at the LHC, currently $M_T \gtrsim 1.3\;\mathrm{TeV}$ (see for example~\cite{ATLAS-CONF-2021-024}). Here we show that the sum of top sector and Gegenbauer potentials can realise viable electroweak breaking without any $v/f$ fine-tuning and with significantly less total tuning than in standard pNGB Higgs scenarios. 

\begin{figure}[t]
\begin{center}
\adjustbox{max width=\textwidth}{%
	\raisebox{8mm}{\includegraphics{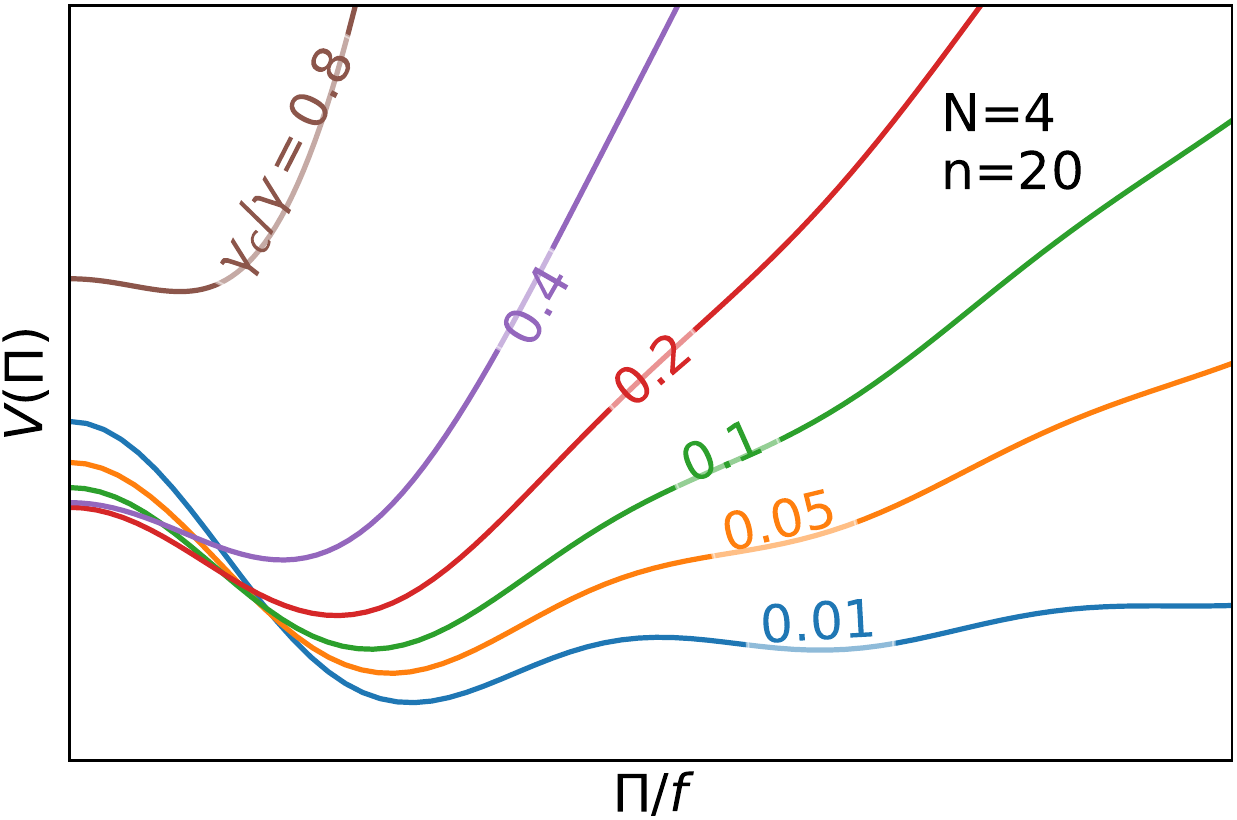}}\qquad
	\includegraphics{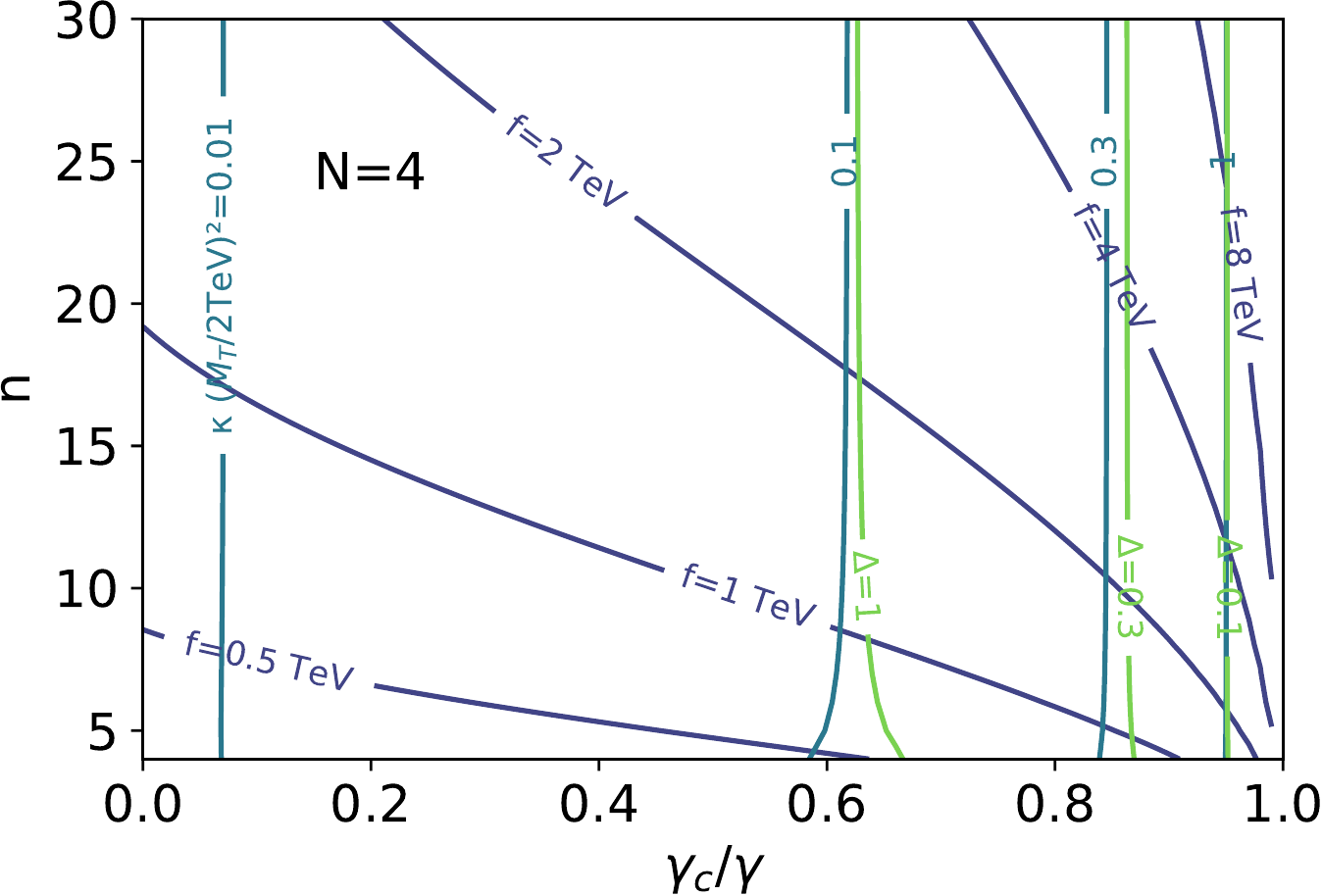}%
}%
\caption{\label{fig:potential} {\it (Left)} Shape of the Gegenbauer-Higgs potential of \autoref{eq:ge-higgs-pot} for $N=4$, $n = 20$ and different values of $\gamma_c/\gamma$.
The global minimum approaches the origin as $\gamma$ tends to its critical value while, for small $\gamma_c/\gamma$, its location is determined by the Gegenbauer contribution to the potential.
{\it (Right)} Parameter space of the Gegenbauer-Higgs model.
Values of the log-derivative $f/v$ tuning are shown as $\Delta$ contours, while $\kappa$ quantifies the tuning of top-sector contributions to the Higgs mass.
The two tunings evolve in opposite ways as functions of $\gamma_c/\gamma$ and both admit values above $10\%$ in the $0.6\,$--$\,0.95$ region for top-partner masses of about $2\,$TeV.
}
\end{center}
\end{figure}

The leading term of the top sector potential is expected to be of the form
\beq\label{eq:V_top}
V_{t} = \kappa \frac{N_c }{16\pi^2} y_t^2 f^2 M_T^2 \sin^2 \GG/f~,
\eeq
where $M_T$ is the lightest top partner mass scale and $\kappa$ is a dimensionless parameter whose size and sign depend on the details of the top sector realisation. Contrary to the standard narrative in pNGB models, where a negative $\sin^2 \GG/f$ term is balanced by a positive $\sin^4 \GG/f$ one to obtain electroweak breaking, here we assume a {\it positive} sign for $V_t$. We will see momentarily that there is no impediment in obtaining this sign with the most common choices of $\mathrm{SO}(N+1)$ representations for the embeddings of $q_L$ and $t_R$. Pressing on, we add a Gegenbauer potential with even $n$ assuming that this arises from an additional source of explicit symmetry breaking in the UV, obtaining the toy potential which is the sum of the two
\beq
V(\GG) = \kappa \frac{N_c }{16\pi^2} y_t^2 f^2 M_T^2 \left[ \sin^2 \GG/f + \gamma\, G_n^{(N-1)/2} (\cos \GG/f) \right]~.
\label{eq:ge-higgs-pot}
\eeq
As illustrated in the left panel of \autoref{fig:potential}, there exists a critical value for the coefficient $\gamma$
\beq
\gamma_{c} = - \frac{(\sin^2 \GG/f)''}{(G_n^{(N-1)/2}(\cos \GG/f))''}\Bigg|_{\GG \,=\, 0}~,
\eeq
below which the global minimum of $V(\Pi)$ is at the origin, whereas for $\gamma > \gamma_c$ the effect of the Gegenbauer is sufficiently strong to ensure $v/f = \sin \langle\GG\rangle/f \neq 0$. This critical value is numerically small, with approximate scaling given by $\gamma_c\approx 8\cdot 10^{-4} (10/n)^{3.6}$, ensuring that the condition $\epsilon n^2 \ll 1$ discussed in~\autoref{sec:CW} is well satisfied.

As shown in the right panel of \autoref{fig:potential}, for very small $\gamma_c/\gamma$, the Gegenbauer contribution to the potential dominates and $f/v \approx n/5.1$ can admit large values without any fine tuning. By contrast, in standard pNGB models $f \gg v$ can only be obtained at the cost of a $\sim v^2/f^2$ cancellation.
Adopting a log-derivative definition,
\begin{equation}
\Delta = \left(\frac{\partial \log f/v}{\partial \log \gamma}\right)^{-1}\:,
\label{eq:vof-tuning}
\end{equation}
one finds here that no tuning is generated from $f/v$ up to $\gamma_c/\gamma \approx 0.6$.
In this region of small $\gamma_c/\gamma$, generating the $125\;\mathrm{GeV}$ Higgs mass while satisfying the LHC bound on coloured top partners however requires to tune the top sector contribution with $\kappa \ll 1$. 

Conversely, for larger $\gamma_c/\gamma$, the Gegenbauer acts as a small perturbation to the top sector potential, shifting the minimum slightly away from the origin and resulting in $f/v \gg n/5.1$ at the price of increasing fine tuning.
The observed Higgs mass is however obtained for $\kappa$ of order one and no tuning is needed in the top sector.

\begin{figure}[t]
\begin{center}
\adjustbox{max width=\textwidth}{%
	\includegraphics[trim=0 0 75 0, clip]{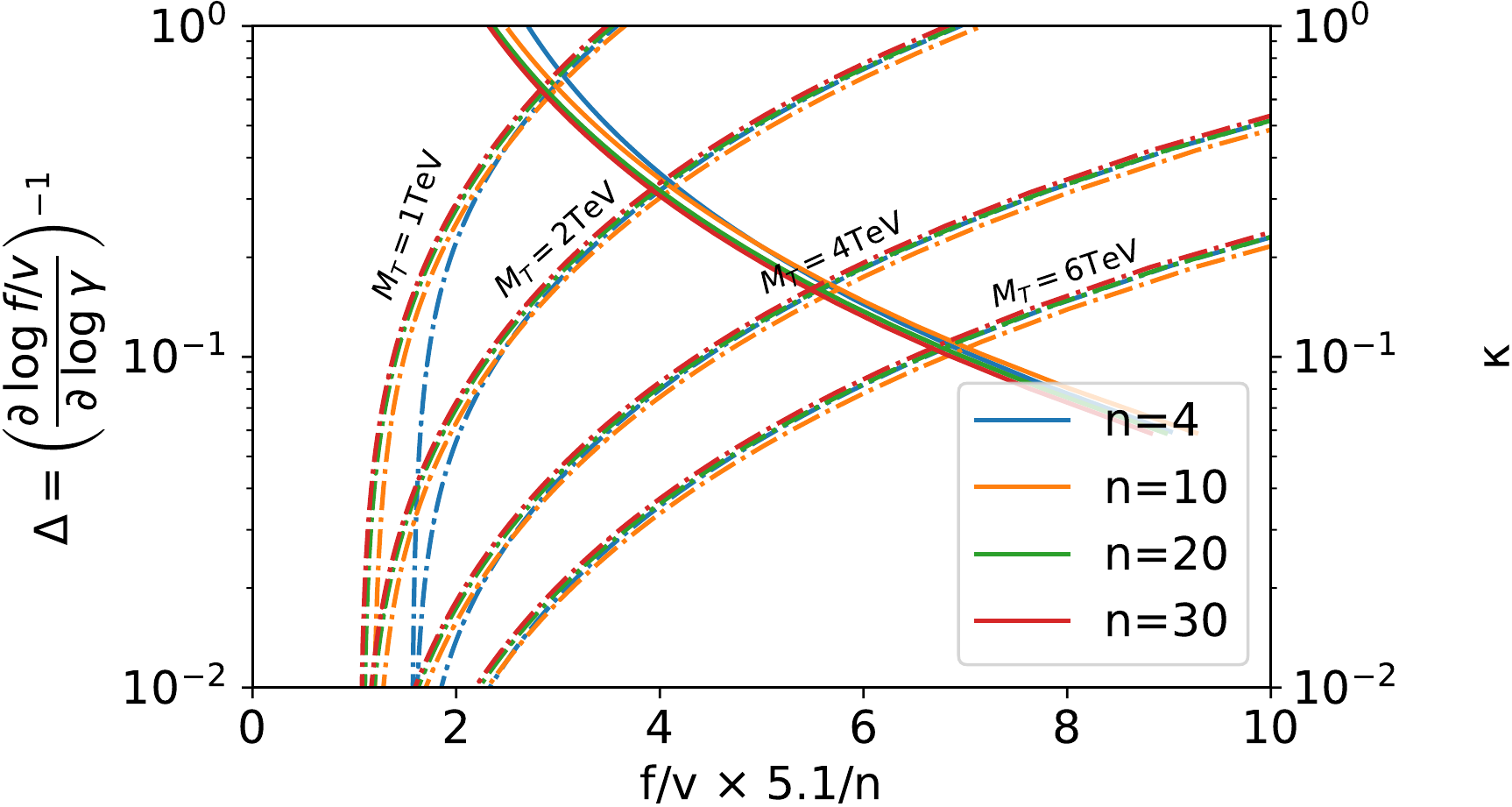}%
	\includegraphics[trim=75 0 0 0, clip]{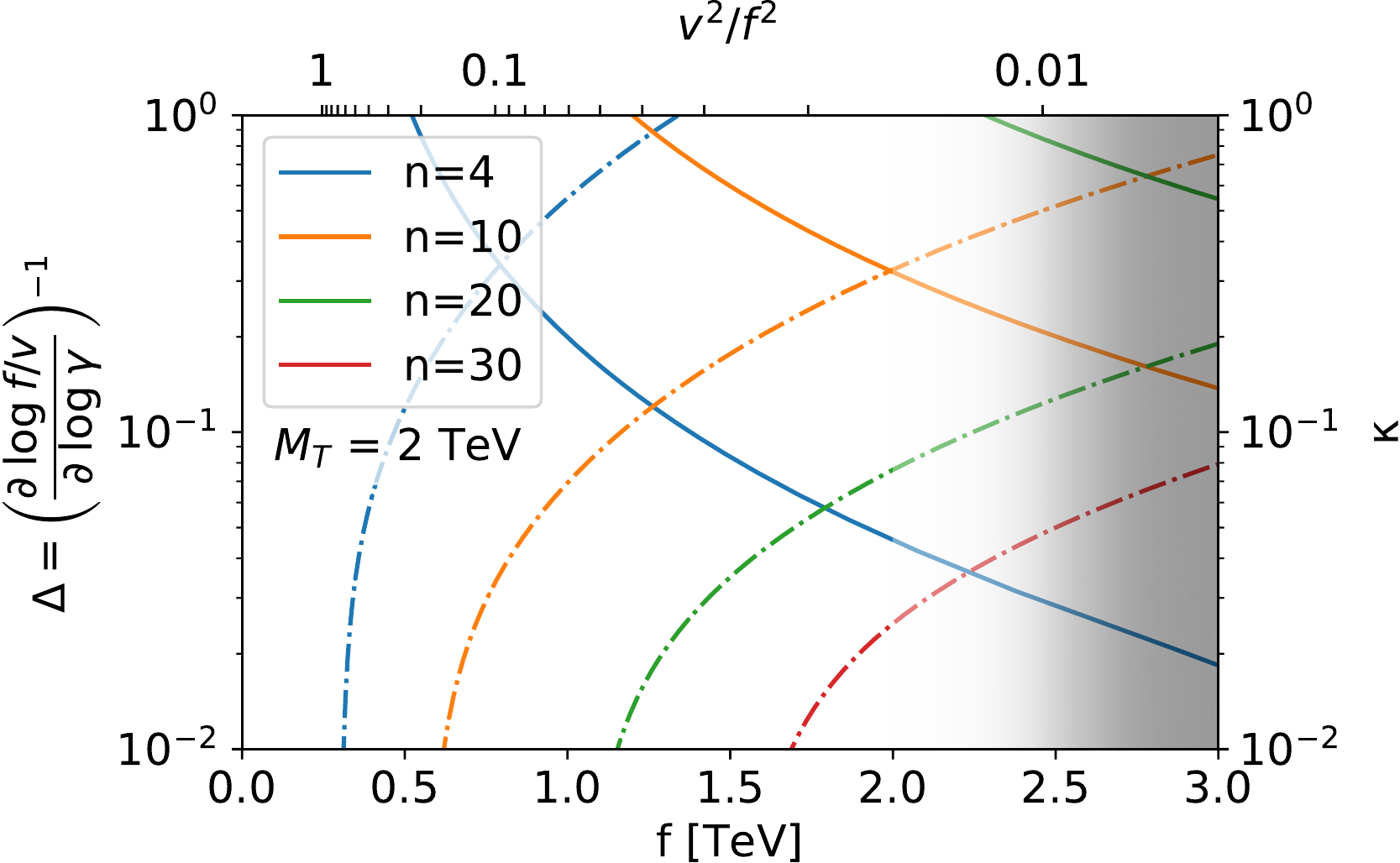}%
}%
\caption{\label{fig:tuning1} Sources of fine tuning in the Gegenbauer Higgs model for $N=4$.
Solid lines provide the log-derivative tuning of $f/v$ that is independent of $M_T$ (left axis), while dashed lines correspond to the Higgs mass tuning arising from the top sector and controlled by $\kappa$ (right axis).
In the right panel, the shaded region for which $f\gtrsim \sqrt{2} M_T/y_t$ is inaccessible in coloured top partner models reproducing the observed top-quark mass.}
\end{center}
\end{figure}

The above features suggest that the most natural parameter space should lie in an intermediate regime where both tunings are present but moderate. This is shown in the left panel of \autoref{fig:tuning1}.
In this regime, the approximate scaling of the two tunings is
\begin{equation}
\Delta \approx 30\%\:\left(\frac{f}{4v} \frac{5.1}{n}\right)^{-2.1}
\:,\qquad
\kappa \approx 30\%\left(\frac{f}{4v} \frac{5.1}{n} \frac{2\,\text{TeV}}{M_T} \right)^{2}
\:.
\end{equation}%
So both reach only about $30\%$ for $M_T \approx 2 \;\mathrm{TeV}$, which roughly corresponds to the ultimate reach of the LHC, and $f \approx 1\,\text{TeV} \times n/5.1$. For such a top partner mass, $n$ cannot be taken much above $10$.
Indeed, in concrete top partner constructions, $M_T$ sets an upper bound on the pNGB decay constant, $f \lesssim \sqrt{2} M_T/y_t$, to reproduce the top-quark mass.\footnote{This is observed in a variety of explicit models. For instance, a mass matrix $$\begin{pmatrix} y f \sin \GG/f & 0 \\  y f \cos \GG/f & M_T \end{pmatrix}$$ among $(\bar{t}_L, \overline{T}_L)$ and $(t_R, T_R)$ yields, expanding for $M_T \ll y f$, the top mass $m_t \approx M_T \sin \GG/f$, resulting in $M_T \gtrsim y_t f/\sqrt{2}$.} This constraint forces the Higgs mass tuning to increase for large $n$ since it implies $\kappa \lesssim 30\%\: (14/n)^2$.
As seen in the right panel of \autoref{fig:tuning1}, values of $n$ below about $10$ are the most natural for $M_T\approx 2\;\mathrm{TeV}$.
For instance, with $n = 10$ and $M_T \approx f \approx 2\;\mathrm{TeV}$ the total tuning is better than $10\%$, when conservatively estimated by multiplying the individual sources.
This is a significant improvement compared to the minimal amount of cancellation necessary in standard pNGB Higgs models with the same $f$, namely $2 v^2/f^2 \approx 3\%$.

We now return to our estimate of the top sector potential in \autoref{eq:V_top} to show that, in general, its sign can be positive, and its size can be tuned by realising $\kappa \ll 1$ through appropriate choices of model parameters.
We illustrate these properties by considering a two-site realisation~\cite{Matsedonskyi:2012ym} of the well-known $\mathbf{5}_L + \mathbf{5}_R$ composite Higgs model~\cite{Contino:2006qr}, where the leading term of the radiatively generated potential is $V_t^{\mathbf{5} + \mathbf{5}} = \alpha_t  \sin^2 \GG / f $ with
\beq
\alpha_t \approx \frac{N_c}{8\pi^2}(y_L^2 - 2 y_R^2) f^2 (M_1^2 - M_4^2) \int \frac{dp \,p^3}{(p^2 + M_1^2 + y_R^2 f^2)(p^2 + M_4^2 + y_L^2 f^2)}~.
\eeq
Evidently, by appropriately choosing the elementary/composite couplings $y_{L,R}$ and the $SO(4)$-preserving composite masses $M_{1,4}\,$, either sign can be obtained for $\alpha_t$ and its size reduced below the natural expectation. The fact that $\alpha_t$ is logarithmically UV divergent is a limitation of the two-site structure, and does not affect our qualitative conclusions.
If we were to consider a three-site model (or phenomenological Weinberg sum rules) to render $\alpha_t$ UV-finite, we would obtain a more elaborate expression, but still taking either sign across the parameter space and with tunable size.\footnote{If both $q_L$ and $t_R$ are partially composite in the $\mathbf{5}_L + \mathbf{5}_R$ model, the parametric scaling of $V_t$ is in fact closer to $y_t f M_T^3$, than to $y_t^2 f^2 M_T^2$ assumed in \autoref{eq:V_top}.
Adopting the former scaling has a mild effect on our results. The $\sim y_t^2 f^2 M_T^2$ dependence is obtained for example in models where the $t_R$ is a fully composite chiral fermion, such as $\mathbf{5}_L + \mathbf{1}_R$ or $\mathbf{14}_L + \mathbf{1}_R$, where $\kappa \ll 1$ can be realised by tuning appropriate combinations of the composite masses.
In the $\mathbf{14}_L + \mathbf{1}_R$ setup a $\sin^4 \GG/f$ term is generated as well at leading order, whose inclusion does not qualitatively change our discussion as long as it has positive sign.}
Finally, we observe that loops from the gauge sector always generate a positive contribution $\alpha_g$, of subleading magnitude compared to top sector loops.
Thus, including the gauge contribution in the potential does not affect our conclusions. 

\subsection{Phenomenology}
The key finding of \autoref{sec:toward_realistic} is that, by including a Gegenbauer potential on top of the one radiatively generated by the top sector, it is possible to obtain natural pNGB Higgs models where the top partner masses are at $M_T \gtrsim 2\;\mathrm{TeV}$ and the correction to the $hVV$ coupling is below $1\%$.
This pushes the two prime observables characteristic of this scenario beyond the sensitivity reach of the HL-LHC.
For instance, for $M_T \approx f \approx 2\;\mathrm{TeV}$ and an explicit breaking spurion in the $n = 10$ index traceless symmetric tensor representation, a conservative estimate places the tuning at better than $10\%$.

\begin{figure}[t]
\begin{center}
\adjustbox{max width=\textwidth}{%
\includegraphics{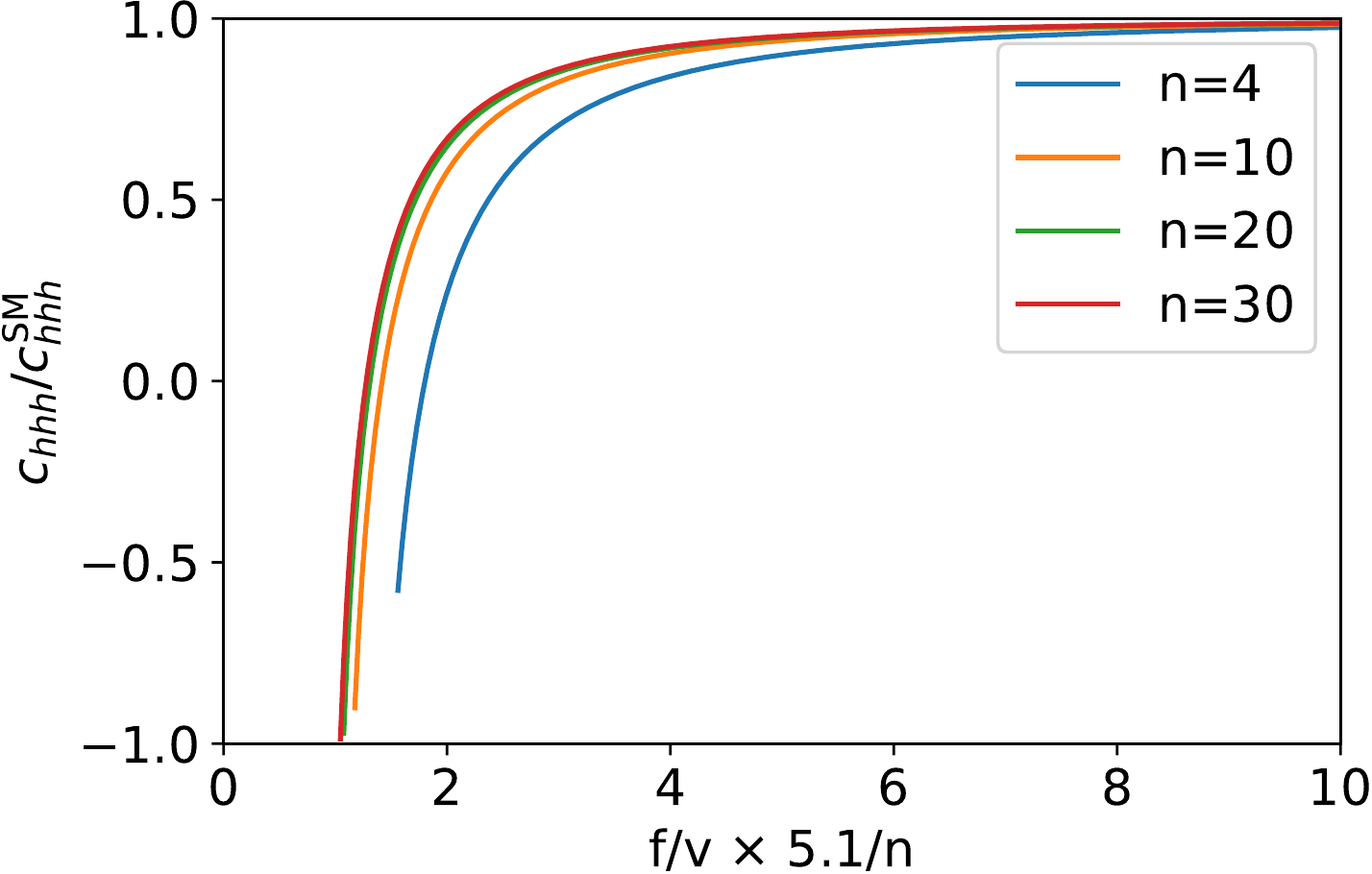}%
\includegraphics[trim=64 0 0 0, clip]{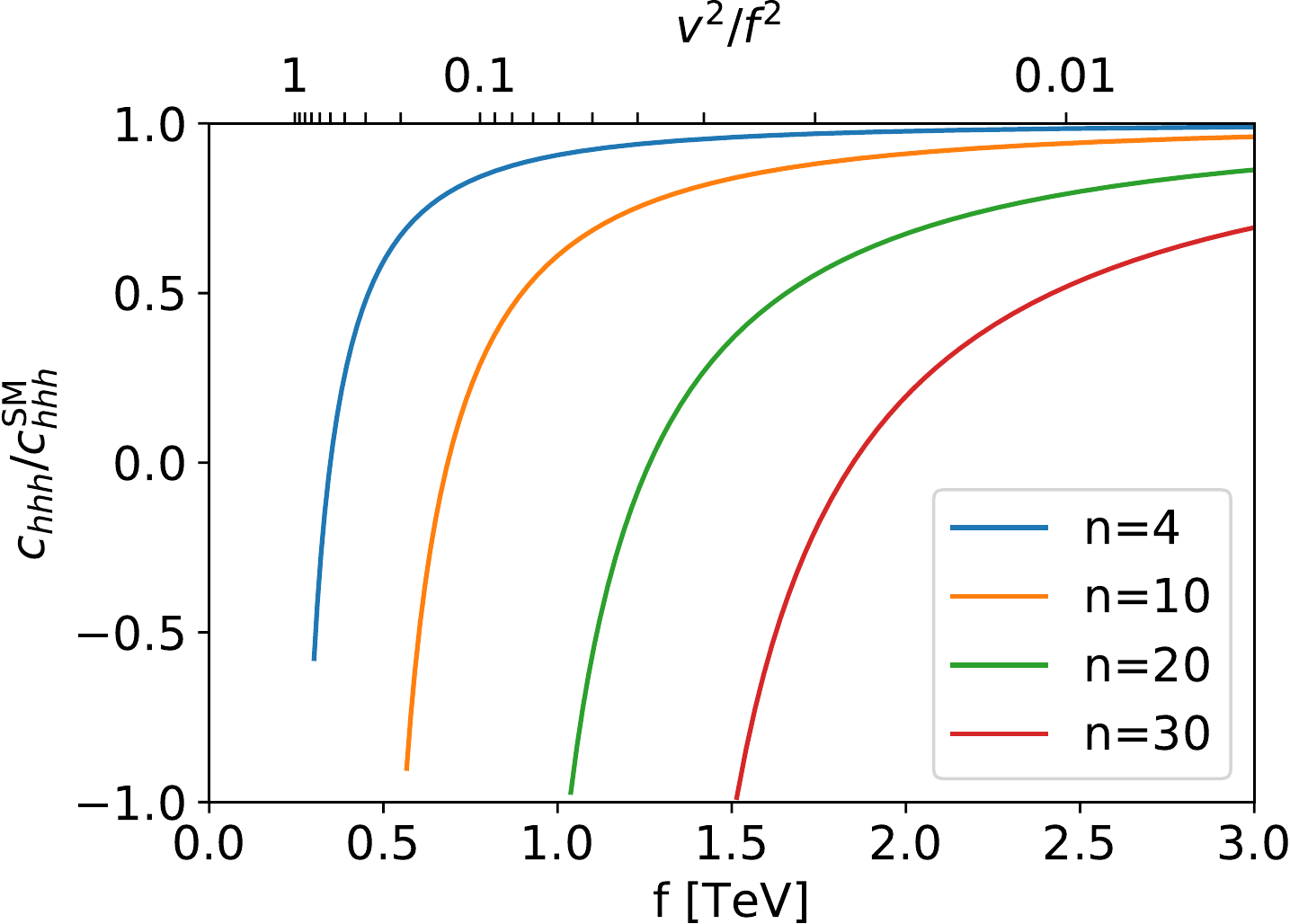}%
}%
\caption{\label{fig:trilinear} The Higgs trilinear coupling normalised to its SM value, for the potential in \autoref{eq:ge-higgs-pot} including both Gegenbauer and top-sector contributions.
Its behaviour as a function of $f$ is independent of the top partner mass scale $M_T$.}
\end{center}
\end{figure}
It is also important to ask how much the Higgs trilinear coupling deviates from the SM in the most natural parameter space, after we include both the Gegenbauer and top sector components. 
In~\autoref{fig:trilinear} we observe the expected limiting behaviours, namely $c_{hhh}/c_{hhh}^{\rm SM}$ is close to $-1$ in the Gegenbauer-dominated region with $f/v \approx n/5.1$, whereas it approaches $+1$ where the Gegenbauer provides only a small perturbation to the top sector potential, resulting in $f/v \gg n/5.1$.
In this latter regime, the following scaling is realised:
\begin{equation}
\frac{c_{hhh}}{c_{hhh}^\text{SM}} \approx 1-1.2\left(\frac{f}{v} \frac{5.1}{n+\lambda}\right)^{-2}
\:.
\end{equation}
For $f$ and $M_T$ of a couple of TeV's, the discussion in the previous section indicated the most natural region has $n$ below about $10$. In this regime, deviations in the trilinear Higgs self coupling are below $10\%$ and out of the HL-LHC reach. Larger modifications would require an increase of the Higgs mass tuning from the top sector. Observable deviations can nevertheless coexist with conceivable tuning levels:
with $n = 30$ and $M_T \approx f\approx 2\;\mathrm{TeV}$ we for example find $c_{hhh}/c_{hhh}^{\rm SM} \approx 0.2$, while the $\kappa$ tuning is about $2\%$.

The above observations suggest that once departures from the oft-assumed minimality of explicit breaking are accepted, testing relatively natural pNGB Higgs models requires an accelerator program capable of reaching beyond the ultimate sensitivity of the LHC.

\section{Conclusions}\label{sec:concl}
The phenomenology of pNGBs is couched firmly in the framework of EFTs. Our present expectations for the `generic' properties of pNGBs are thus guided by EFT logic in combination with lessons learned from commonly studied models.
It is important to identify which of these expectations are driven primarily by EFT reasoning and which are driven instead by the assumed structure of the models.  One commonly injected element is minimality.
Not minimality in the sense of the number of degrees of freedom, or the rank of the broken symmetry, but in terms of the representations of spurions that explicitly break the symmetries and generate the vacuum potential of pNGBs.

In this work, this question has been examined from the perspective of the radiative stability of pNGB potentials.
This is a useful guiding principle: if a pNGB potential is not radiatively stable against UV corrections, one cannot develop any robust expectations for its `generic' properties.
For $\SONp\to\SON$, we have found that Gegenbauer polynomials form a basis of radiatively stable pNGB scalar potentials.
As a result, there exist universality classes of natural microscopic theories which flow, in the IR, to a theory with pNGBs whose scalar potential is predominantly of the form of a Gegenbauer polynomial of degree $n$, up to corrections at second order in some small parameter.
This is an internal-symmetry analogue of the usual multipole expansion in momentum-space.

For small $n$, the Gegenbauer polynomials furnish the usual sine and cosine-type potentials found in typical symmetry breaking scenarios, reflecting a `minimal' spurion.
However for large $n$ the vacuum structure differs markedly.
In particular, the location of the global minimum and pNGB expectation value `$v$' may be at arbitrarily small values, since $v \propto f/n$.
Furthermore, the potential locally becomes approximately periodic, with period $2 \pi f/n$.
This is interesting as it appears to be, at an approximate level, akin to the potential for an Abelian pNGB with explicit breaking via a spurion of charge $n$, except that in our case the pNGBs are non-Abelian.

We conclude that typical expectations for a non-Abelian pNGB potential are driven not by any fundamental aspect of EFT itself but, instead, by the expectation that low-dimensional representations for explicit-breaking spurions are more generic than higher-dimension ones. This may indeed be the case in many realistic settings, since often the leading breaking of global symmetries is via a gauge or Yukawa interaction and both typically correspond to a low-dimension spurion.  As a result, we do not question the genericity of this expectation but emphasise that there are additional technically natural possibilities.

There are a number of questions that remain unanswered.
One question concerns different symmetry groups.
The Gegenbauer polynomials arose due to the assumption of an $\SON$ symmetry, but for different symmetry groups it is not clear what the associated functions would be.
Another question concerns the role of the UV symmetry.
Throughout, for the symmetry breaking pattern $\mathcal{G}\to\mathcal{H}$, we made a specific assumption for $\mathcal{G}$.
On the other hand, it has become evident that certain properties of pNGBs may be determined independently of any specific choice for $\mathcal{G}$ \cite{Low:2014nga}.
It would be interesting to know if this is the case for the extraction of the radiatively stable pNGB potentials.

On a more phenomenological note, it is important to understand if the Higgs boson could be a pNGB.
To this end we have developed, at a relatively superficial level, a class of `Gegenbauer Higgs' models.
These models adopt the standard structure for a composite pNGB Higgs, with the additional ingredient of an extra source of explicit symmetry breaking in the UV which is in a high-dimension traceless symmetric irrep of a global $\text{SO}(5)$ symmetry.\footnote{Note that the SM pion potential is shaped by two explicit symmetry breaking contributions of independent origin: one internal to the strong sector (the quark masses) and one arising from the couplings of the external gauge fields (primarily the photon). However, in that case the explicit symmetry breaking is in low dimensional representations.}
This gives rise to an additional Gegenbauer polynomial potential in the IR which is a radiatively stable and thus a technically natural IR augmentation of the usual models.  This greatly modifies the vacuum structure of pNGB Higgs models.

Primarily, the Gegenbauer potential can create a global minimum at small field values, hence a natural hierarchy $v\ll f$ is realised.
Since Higgs coupling modifications scale proportional to $v^2/f^2$, this allows for a pNGB Higgs to be naturally SM-like.
On the contrary, the explicit symmetry breaking in the top and gauge sectors remains.
Thus the Gegenbauer Higgs class of models is not immune to fine-tuning arising from the lack of complete $\text{SO}(5)$ multiplets in these sectors at the electroweak scale.
This becomes the dominant source of fine-tuning and we find that for all new coloured resonances above about $2\,$TeV some combined fine-tuning remains at the $\mathcal{O}(10\%)$ level.
This ultimately suggests that, if we are to truly understand whether or not the Higgs is a composite pNGB, then we will need to probe very small Higgs coupling modifications or push coloured resonance searches to energies beyond the reach of the HL-LHC.

There are a number of open questions concerning the Higgs models.
Presumably the most pressing is to identify if there are UV possibilities that could motivate such a high-dimensional symmetry-breaking spurion.
Perhaps some alternative interpretation or motivation would be found by appealing to a holographic picture?
Furthermore, the phenomenology of these models has also not been explored to any extent.
For instance, perhaps such highly oscillating scalar potentials could have a cosmological impact relating to the nature of the electroweak phase transition (see e.g.~\cite{DiLuzio:2019wsw}).
On the model-building side, it would be interesting to explore different symmetry groups or, more importantly, whether the symmetry structure of the top sector could be modified such that the generated contributions to the effective potential were also of a Gegenbauer form.
If possible, this may significantly reduce fine-tuning.\footnote{It would be interesting if, for instance, some variant of the $\mathcal{Z}_n$ models obtained in \cite{Hook:2018jle} may be embedded within the non-Abelian structures found here.} Finally, the approach developed here could possibly be applied to neutral naturalness setups such as the Twin Higgs framework~\cite{Chacko:2005pe}. The reason being that the $v/f$ tuning in this framework is often the dominant one whereas the top partners, being uncoloured, can in principle be quite light. It would seem that combining this with the properties of the Gegenbauer Higgs setup may be fruitful.

\section*{Acknowledgments}
We thank Csaba Cs\'{a}ki and Gian Giudice for comments on the manuscript and Henriette Elvang and Aaron Pierce for discussions.

\appendix
\section{Large-\texorpdfstring{$n$}{n} Limit}
\label{app:large-n}
A large-$n$ approximation for the Gegenbauer potential can be obtained by considering the differential equation satisfied by Gegenbauer polynomials $G^\lambda_n(x)$:
\begin{equation}
(1-x^2)G'' - (2\lambda+1) x G' + n(n+2\lambda)G=0\,.
\end{equation}
Performing the $x\to\cos\frac{y}{n+\lambda}$ change of variable, it becomes:
\begin{equation}
(n+\lambda)^2 G'' + 2\lambda(n+\lambda)\cot\frac{y}{n+\lambda} G' + n(n+2\lambda) G = 0\,,
\end{equation}
which expands for large-$n$ to
\begin{equation}
(G''+G)y+2\lambda G' \simeq 0\,,
\end{equation}
up to $1/n^2$ corrections (while the change of variable $x\to \cos \frac{y}{n}$ would have left out $1/n$
corrections).
Imposing that its solution goes to $1$ for $y=0$ and $\lambda=1/2$, as Legendre polynomials do, one finds a Bessel function of the first kind.
In the large argument limit, the latter also approximate to a cosine.
Thus, as stated in \autoref{eq:geg-approx},
\begin{equation}
G^\lambda_n\left(\cos\frac{y}{n+\lambda}\right)
	\;\xrightarrow{n\gg1}\; \frac{J_{\lambda-1/2}(y)}{y^{\lambda-1/2}}
	\;\xrightarrow{y\gg1}\; \sqrt{\frac{2}{\pi}} \frac{\cos\left(y-\lambda\frac{\pi}{2}\right)}{y^\lambda}
	\,.
\end{equation}

\bibliographystyle{JHEP}
\bibliography{biblio}

\providecommand{\href}[2]{#2}\begingroup\raggedright\begin{thebibliography}{10}

\bibitem{Goldstone:1961eq}
J.~Goldstone, \emph{{Field Theories with Superconductor Solutions}},
  \href{https://doi.org/10.1007/BF02812722}{\emph{Nuovo Cim.} {\bfseries 19}
  (1961) 154--164}.

\bibitem{Nambu:1960tm}
Y.~Nambu, \emph{{Quasiparticles and Gauge Invariance in the Theory of
  Superconductivity}},
  \href{https://doi.org/10.1103/PhysRev.117.648}{\emph{Phys. Rev.} {\bfseries
  117} (1960) 648--663}.

\bibitem{Coleman:1973jx}
S.~R. Coleman and E.~J. Weinberg, \emph{{Radiative Corrections as the Origin of
  Spontaneous Symmetry Breaking}},
  \href{https://doi.org/10.1103/PhysRevD.7.1888}{\emph{Phys. Rev. D} {\bfseries
  7} (1973) 1888--1910}.

\bibitem{Brezin:1976ap}
E.~Br\'ezin, J.~Zinn-Justin and J.~C. Le~Guillou, \emph{{Renormalization of the
  nonlinear $\sigma$ model in $2+\epsilon$ dimensions}},
  \href{https://doi.org/10.1103/PhysRevD.14.2615}{\emph{Phys. Rev. D}
  {\bfseries 14} (1976) 2615}.

\bibitem{Contino:2010rs}
R.~Contino, \emph{{The Higgs as a Composite Nambu-Goldstone Boson}},  in
  \emph{{Theoretical Advanced Study Institute in Elementary Particle Physics}:
  {Physics of the Large and the Small}}, pp.~235--306, 2011,
  \href{https://arxiv.org/abs/1005.4269}{{\ttfamily 1005.4269}},
  \href{https://doi.org/10.1142/9789814327183_0005}{DOI}.

\bibitem{Bellazzini:2014yua}
B.~Bellazzini, C.~Cs\'aki and J.~Serra, \emph{{Composite Higgses}},
  \href{https://doi.org/10.1140/epjc/s10052-014-2766-x}{\emph{Eur. Phys. J. C}
  {\bfseries 74} (2014) 2766},
  [\href{https://arxiv.org/abs/1401.2457}{{\ttfamily 1401.2457}}].

\bibitem{Panico:2015jxa}
G.~Panico and A.~Wulzer, \emph{{The Composite Nambu-Goldstone Higgs}},
  vol.~913.
\newblock Springer, 2016,
  \href{https://doi.org/10.1007/978-3-319-22617-0}{10.1007/978-3-319-22617-0}.

\bibitem{Harnik:2016koz}
R.~Harnik, K.~Howe and J.~Kearney, \emph{{Tadpole-Induced Electroweak Symmetry
  Breaking and pNGB Higgs Models}},
  \href{https://doi.org/10.1007/JHEP03(2017)111}{\emph{JHEP} {\bfseries 03}
  (2017) 111}, [\href{https://arxiv.org/abs/1603.03772}{{\ttfamily
  1603.03772}}].

\bibitem{Galloway:2016fuo}
J.~Galloway, A.~L. Kagan and A.~Martin, \emph{{A UV complete partially
  composite-pNGB Higgs}},
  \href{https://doi.org/10.1103/PhysRevD.95.035038}{\emph{Phys. Rev. D}
  {\bfseries 95} (2017) 035038},
  [\href{https://arxiv.org/abs/1609.05883}{{\ttfamily 1609.05883}}].

\bibitem{Csaki:2017cep}
C.~Cs\'aki, T.~Ma and J.~Shu, \emph{{Maximally Symmetric Composite Higgs
  Models}}, \href{https://doi.org/10.1103/PhysRevLett.119.131803}{\emph{Phys.
  Rev. Lett.} {\bfseries 119} (2017) 131803},
  [\href{https://arxiv.org/abs/1702.00405}{{\ttfamily 1702.00405}}].

\bibitem{Coleman:1969sm}
S.~R. Coleman, J.~Wess and B.~Zumino, \emph{{Structure of phenomenological
  Lagrangians. 1.}},
  \href{https://doi.org/10.1103/PhysRev.177.2239}{\emph{Phys. Rev.} {\bfseries
  177} (1969) 2239--2247}.

\bibitem{Callan:1969sn}
C.~G. Callan, Jr., S.~R. Coleman, J.~Wess and B.~Zumino, \emph{{Structure of
  phenomenological Lagrangians. 2.}},
  \href{https://doi.org/10.1103/PhysRev.177.2247}{\emph{Phys. Rev.} {\bfseries
  177} (1969) 2247--2250}.

\bibitem{Alonso:2015fsp}
R.~Alonso, E.~E. Jenkins and A.~V. Manohar, \emph{{A Geometric Formulation of
  Higgs Effective Field Theory: Measuring the Curvature of Scalar Field
  Space}}, \href{https://doi.org/10.1016/j.physletb.2016.01.041}{\emph{Phys.
  Lett. B} {\bfseries 754} (2016) 335--342},
  [\href{https://arxiv.org/abs/1511.00724}{{\ttfamily 1511.00724}}].

\bibitem{Chetyrkin:1980pr}
K.~G. Chetyrkin, A.~L. Kataev and F.~V. Tkachov, \emph{{New Approach to
  Evaluation of Multiloop Feynman Integrals: The Gegenbauer Polynomial x Space
  Technique}}, \href{https://doi.org/10.1016/0550-3213(80)90289-8}{\emph{Nucl.
  Phys. B} {\bfseries 174} (1980) 345--377}.

\bibitem{Kazakov:1986mu}
D.~I. Kazakov and A.~V. Kotikov, \emph{{The Method of Uniqueness: Multiloop
  Calculations in {QCD}}},
  \href{https://doi.org/10.1007/BF01041909}{\emph{Teor. Mat. Fiz.} {\bfseries
  73} (1987) 348--361}.

\bibitem{Agashe:2004rs}
K.~Agashe, R.~Contino and A.~Pomarol, \emph{{The Minimal Composite Higgs
  model}}, \href{https://doi.org/10.1016/j.nuclphysb.2005.04.035}{\emph{Nucl.
  Phys. B} {\bfseries 719} (2005) 165--187},
  [\href{https://arxiv.org/abs/hep-ph/0412089}{{\ttfamily hep-ph/0412089}}].

\bibitem{ATLAS-CONF-2021-024}
{{\rm ATLAS collaboration}}, \emph{{Search for pair-production of vector-like
  quarks in $pp$ collision events at $\sqrt{s}\,$=$\,$13~TeV with at least one
  leptonically-decaying $Z$~boson and a third-generation quark with the ATLAS
  detector}},  \href{https://cds.cern.ch/record/2773300}{ATLAS-CONF-2021-024}.

\bibitem{Matsedonskyi:2012ym}
O.~Matsedonskyi, G.~Panico and A.~Wulzer, \emph{{Light Top Partners for a Light
  Composite Higgs}}, \href{https://doi.org/10.1007/JHEP01(2013)164}{\emph{JHEP}
  {\bfseries 01} (2013) 164},
  [\href{https://arxiv.org/abs/1204.6333}{{\ttfamily 1204.6333}}].

\bibitem{Contino:2006qr}
R.~Contino, L.~Da~Rold and A.~Pomarol, \emph{{Light custodians in natural
  composite Higgs models}},
  \href{https://doi.org/10.1103/PhysRevD.75.055014}{\emph{Phys. Rev. D}
  {\bfseries 75} (2007) 055014},
  [\href{https://arxiv.org/abs/hep-ph/0612048}{{\ttfamily hep-ph/0612048}}].

\bibitem{Low:2014nga}
I.~Low, \emph{{Adler\textquoteright{}s zero and effective Lagrangians for
  nonlinearly realized symmetry}},
  \href{https://doi.org/10.1103/PhysRevD.91.105017}{\emph{Phys. Rev. D}
  {\bfseries 91} (2015) 105017},
  [\href{https://arxiv.org/abs/1412.2145}{{\ttfamily 1412.2145}}].

\bibitem{DiLuzio:2019wsw}
L.~Di~Luzio, M.~Redi, A.~Strumia and D.~Teresi, \emph{{Coset Cosmology}},
  \href{https://doi.org/10.1007/JHEP06(2019)110}{\emph{JHEP} {\bfseries 06}
  (2019) 110}, [\href{https://arxiv.org/abs/1902.05933}{{\ttfamily
  1902.05933}}].

\bibitem{Hook:2018jle}
A.~Hook, \emph{{Solving the Hierarchy Problem Discretely}},
  \href{https://doi.org/10.1103/PhysRevLett.120.261802}{\emph{Phys. Rev. Lett.}
  {\bfseries 120} (2018) 261802},
  [\href{https://arxiv.org/abs/1802.10093}{{\ttfamily 1802.10093}}].

\bibitem{Chacko:2005pe}
Z.~Chacko, H.-S. Goh and R.~Harnik, \emph{{The Twin Higgs: Natural electroweak
  breaking from mirror symmetry}},
  \href{https://doi.org/10.1103/PhysRevLett.96.231802}{\emph{Phys. Rev. Lett.}
  {\bfseries 96} (2006) 231802},
  [\href{https://arxiv.org/abs/hep-ph/0506256}{{\ttfamily hep-ph/0506256}}].

\end{thebibliography}\endgroup

\end{document}